\documentclass[]{elsart}
\usepackage{graphicx}
\usepackage{amssymb}
 
\begin{document}
\intextsep 30pt
\textfloatsep 30pt
\begin{frontmatter}
\title{
Direction-sensitive dark matter search results in a surface laboratory}
\author{Kentaro~Miuchi$^a$},
\ead{miuchi@cr.scphys.kyoto-u.ac.jp}
\author{Kaori~Hattori$^a$}, 
\author{Shigeto~Kabuki$^a$}, 
\author{Hidetoshi~Kubo$^a$},
\author{Shunsuke~Kurosawa$^a$}, 
\author{Hironobu~Nishimura$^a$}, 
\author{Yoko~Okada$^a$}, 
\author{Atsushi~Takada$^a$},
\author{Toru~Tanimori$^a$},
\author{Ken'ichi~Tsuchiya$^a$}, 
\author{Kazuki~Ueno$^a$}, 
\author{Hiroyuki~Sekiya$^b$}, 
\author{Atsushi~Takeda$^b$}

\address{$^a$ Cosmic-Ray Group, Department of Physics, Graduate School of Science, 
Kyoto University Kitashirakawa-oiwakecho, Sakyo-ku, Kyoto, 606-8502, Japan}
\address{$^{b}$  Kamioka Observatory, ICRR, The University of Tokyo
Higashi-Mozumi, Kamioka cho, Hida 506-1205 Japan}

\begin{abstract}

We developed a three-dimensional gaseous tracking device 
and performed a direction-sensitive 
dark matter search in a surface laboratory.
By using 150 Torr carbon-tetrafluoride ($\rm CF_4$) gas, 
we obtained a sky map drawn with 
the recoil directions of the carbon and fluorine nuclei,
and set the first limit on the spin-dependent WIMP (Weakly Interacting 
Massive Particles)-proton cross section 
by a direction-sensitive method.
Thus, we showed that  
a WIMP-search experiment with a gaseous tracking device 
can actually set limits. 
Furthermore, we demonstrated that this method will potentially play a 
certain role in revealing the nature of dark matter 
when a low-background large-volume detector is developed. 
\end{abstract}

\end{frontmatter}
\newpage
\section{Introduction}
\label{section:intro}
Revealing the nature of dark matter has received 
much more attention since 
the successive results of the WMAP 
cosmic microwave background all-sky observation\cite{ref:WMAP Spergel}.
Many dark matter direct search experiments
have been concluded\cite{ref:DAMA_aniso,ref:DAMA_Xe,ref:DAMA_annual,ref:NAIAD,ref:Sekiya_IDM2004,ref:Tokyo_CaF2}, 
are currently being performed\cite{ref:CRESST,ref:ZEPLIN,ref:SIMPLE2005,ref:PICASSO,ref:CDMS2,ref:EDELWEISS,ref:DRIFT2_NIM}, 
or are in the planning stage\cite{ref:superCDMS}.
Weakly interacting massive particles (WIMPs) are very
plausible candidates of dark matter. 
Most dark matter search experiments are designed to measure 
only the energy 
transferred to the nucleus 
through WIMP-nucleus scatterings.
In these experiments, the distinct signals of WIMPs 
are the 
annual modulation of the energy spectrum.
the halo is isotropic.
Because the amplitude of an annual modulation signal is 
only a few~$\%$ of the rate, 
positive signatures of WIMPs are 
very difficult to detect with these detectors\cite{ref:DAMA_annual}.
On the other hand, 
the motion of the solar system with respect to the galactic halo is considered 
to provide a much larger asymmetry in the  direction-distribution 
of the WIMP velocity observed at Earth\cite{ref:Spergel_WIMP}.
A detection method of a positive signature of WIMPs
by measuring the directions of the nuclear recoils have been 
studied\cite{ref:DAMA_aniso,ref:Sekiya_IDM2004,ref:DRIFT2_NIM,ref:Anne}.
If the distribution of the local halo were not isotropic -- 
perhaps quite clumpy--
direction-sensitive methods would then 
potentially provide information 
on it.

Gaseous detectors are one of the most appropriate 
devices to detect this ''WIMP-wind''\cite{ref:Gerbier,ref:Masek}. 
The DRIFT project has performed underground measurements 
for more than two years 
with 1 m$^3$ time projection chambers (TPC)
filled with a low-pressure $\rm CS_2$ gas aiming to detect the 
mainly WIMP-wind via spin-independent(SI) interactions\cite{ref:DRIFT2_NIM}.
Since former studies suggest to 
perform the WIMP-search both via SI and spin-dependent(SD) 
interactions\cite{ref:SDCollor,ref:SDtheory}, 
we proposed a SD-sensitive WIMP-wind search, named 
NEWAGE (NEw generation WIMP-search
with an Advanced Gaseous tracking device Experiment)\cite{ref:Miuchi_PLB}.  
We developed a $\mu$-TPC with a detection volume of 
23 $\times$ 28  $\times$ 30 $\rm cm^3$ and 
performed a direction-sensitive WIMP-search experiment in a 
surface laboratory.

\section{Detector performance}
\label{section:micto-TPC}
In this section, we describe the detector performance, 
which is 
also described in Ref. \cite{ref:Miuchi_IDM2006}. 
We mainly measured the detector response in the 
energy range that we used for the WIMP-search analysis
(100-400keV, DM energy range). 
The lower threshold (100keV) was set at 
the lowest energy for an effective 
gamma-ray rejection with a 150 Torr $\rm CF_4$ gas,
and the higher threshold (400keV) was set at 
the expected highest recoil energy by the WIMPs at the escape velocity.

\subsection{$\mu$-TPC}
A $\mu$-TPC is a gaseous time projection chamber with a micro pixel chamber 
($\mu$-PIC)
readout. 
A $\mu$-PIC is a gaseous two-dimensional position-sensitive detector 
manufactured by printed circuit board (PCB) technology. 
PCB technology realizes an economical  mass production, 
which is one of the most important 
requirements to fabricate a dark matter detector.
After a general-purposed detector R\&D,
we studied the performance of a $\mu$-TPC with a detection volume of 
$\rm10\times10\times10 cm^3$ 
using a 150 Torr $\rm CF_4$ gas
as our first step towards a dark matter experiment\cite{ref:Sekiya_PSD}. 
In this study, we demonstrated the 
tracking of nuclear recoils and the gamma-ray rejection
with a $\rm CF_4$ gas of 150 Torr.
We also found that we need some more studies 
to operate the $\mu$-TPC with a $\rm CF_4$ gas at a design value of 30 Torr
for more than a few days.
We then developed a large-sized (31 $\times$ 31 $\rm cm^2$) 
$\mu$-PICs\cite{ref:uPIC30}, and subsequently 
developed a large-volume $\mu$-TPC with a detection volume of 
23 $\times$ 28  $\times$ 30 $\rm cm^3$. 
We studied the fundamental properties 
of this large-volume $\mu$-TPC 
with a standard Ar-$\rm C_2 H_6$ gas mixture 
at normal pressure\cite{ref:Miuchi_IWORID8}.
The size is of the $\mu$-PIC is restricted by the working size of the PCB, 
and we plan to realize a larger readout area by placing many pieces, 
for instance 4 $\rm \times$ 4 pieces for a 1 m${}^2$,
of $\mu$-PICs.  
Before we take the next technology steps, we performed a 
first direction-sensitive spin-dependent dark matter search experiment 
to demonstrate that a gaseous detector and a direction-sensitive 
method can actually set limits on the WIMPs. 
We report on the results of 
a direction-sensitive dark matter search experiment 
using this large-volume $\mu$-TPC with a 150 Torr $\rm CF_4$ gas 
(9.0g effective mass) in a surface laboratory.

\begin{figure}[h]
  \begin{center}
    \includegraphics[width=0.9\linewidth]{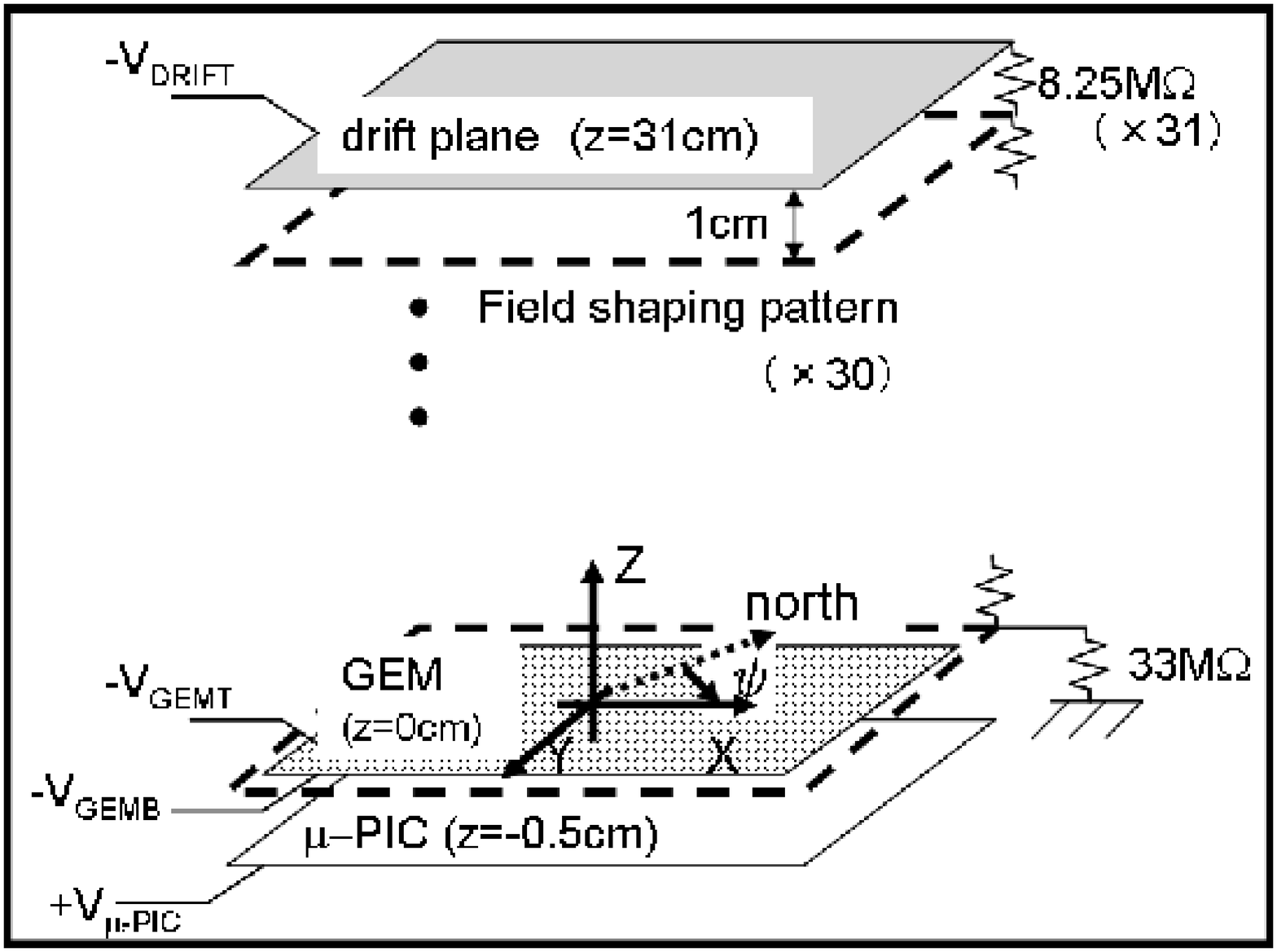}
    \caption{Schematic drawings of the $\mu$-TPC. $\mu$-PIC and the GEM 
are set horizontally.}
    \label{fig:TPC}
  \end{center}
\end{figure}

We used the $\mu$-TPC described in Ref. \cite{ref:Miuchi_IWORID8}, and thus 
most parts of the detector were unchanged.
Herein, we describe only the significant properties and the modified parts 
of the detector system. 
Figure \ref{fig:TPC} is a schematic drawing of the $\mu$-TPC.
A $\mu$-PIC with a detection area of 31 $\times$ 31 $\rm cm^2$
(TOSHIBA/DNP, SN060222-3) was used as  
the main gas-multiplier and as readout electrodes.
A gas electron multiplier (GEM\cite{ref:GEM}) with an effective area of 
23 $\times$ 28 $\rm cm^2$,
which was manufactured by a Japanese company (Scienergy Co. Ltd.), 
was used as a pre-amplifier.
The drift length was 31 cm and 
the detection volume was 
23 $\times$ 28 $\times$ 31 $\rm cm^3$.
The electric field of the drift volume was formed with a 
field-shaping pattern on a fluorocarbon circuit board surrounding 
the detection volume. 
The detector was placed in a 3 mm-thick stainless-steel vacuum chamber 
filled with 150 Torr $\rm CF_4$.
The data-acquisition system, which is described in Ref. \cite{ref:Kubo_IEEE},  
was triggered 
by the signals of the $\mu$-TPC itself.
Hence, the 
absolute Z position was not measured, but only 
the relative positions.
Typical operation parameters were 
$\rm V_{\mu-PIC}=490V$, $\rm V_{GEMB}=630V$, $\rm V_{GEMT}=915V$, and 
$\rm V_{DRIFT}$=8.25kV. $\rm V_{\mu-PIC}$ is the voltage 
supplied to the $\mu$-PIC, 
$\rm V_{GEMT}$ is the voltage supplied to the top of the GEM, 
$\rm V_{GEMB}$ is the voltage supplied to the bottom of the GEM, 
and $\rm V_{drift}$ is the voltage supplied to the drift plane.
These parameters were optimized to realize stable operation
at a combined ($\mu$-PIC $\times$ GEM) gas gain of 2300, which is high 
enough to maintain the detection efficiency of the recoil nucleus.
We still suffer from discharge problems when operating a $\mu$-PIC alone.

\subsection{Energy calibration and energy resolution}
We calibrated and monitored the energy with alpha particles
generated by the $\rm ^{10}B (n,\alpha)^7Li$ (Q=2.7MeV) reaction
\cite{ref:Miuchi_IDM2006}.
We set a glass plate coated with a thin 0.6 $\mu$m $\rm ^{10}B$ layer 
in the $\mu$-TPC, and irradiated the $\mu$-TPC with thermalized neutrons.
Alpha particles (5.6~MeV, 6.1~MeV, and 7.2~MeV) 
from the decays of the radon progeny 
were also used.
The track-length and the energy-deposition correlation down to the  DM
energy range agreed with a calculation by SRIM\cite{ref:SRIM}.

The measured energy resolution in 5--8MeV 
was 50$\%$(FWHM), which was dominated by the gain inhomogeneity 
of the $\mu$-PIC. 
This would be the worst value when the gain inhomogeneity 
of the $\mu$-PIC dominates the energy resolution.
The energy resolution could be dominated by the 
the statistics of the primary electrons.
In this case, 
the energy resolution at 100~keV with a CF$_4$ gas was estimated
from a measurement with an Ar-$\rm C_2H_6$ gas mixture,
because it 
is not very easily measured due to
a small photo-absorption cross-section and a long range of the electron. 
The energy resolution at 60~keV, measured  
with an Ar-$\rm C_2H_6$ gas mixture, was 60$\%$ (FWHM)\cite{ref:Miuchi_IWORID8}.
We caliculataed with 
$\Delta E_{\rm CF_4}=\Delta E_{\rm Ar}\cdot (\frac{60/W_{\rm Ar}}{100/W_{\rm CF_4}})^{1/2}$. 
Here, $\Delta E_{\rm CF_4}$ and $\Delta E_{\rm Ar}$ are 
the energy resolution at 100~keV with a $\rm CF_4$ gas 
and that at 60~keV with an Ar-$\rm C_2H_6$ gas mixture, respectively.
$\rm W_{\rm CF_4}$=54eV  and $\rm W_{Ar}$=26eV are the 
energies needed to created 
an ion-electron pair in a $\rm CF_4$ gas and 
an Ar-$\rm C_2H_6$ gas mixture, respectively.
Thus, the estimated energy resolution was 70$\%$ (FWHM) 
at 100 keV with a CF$_4$ gas, 
if the statistics of the primary electron dominates the energy resolution.
The energy resolution in the DM energy range was therefore estimated to be 
better than 70$\%$.

\subsection{Nuclear recoil detection efficiency}
\begin{figure}[h]
  \begin{center}
    \includegraphics[width=0.48\linewidth]{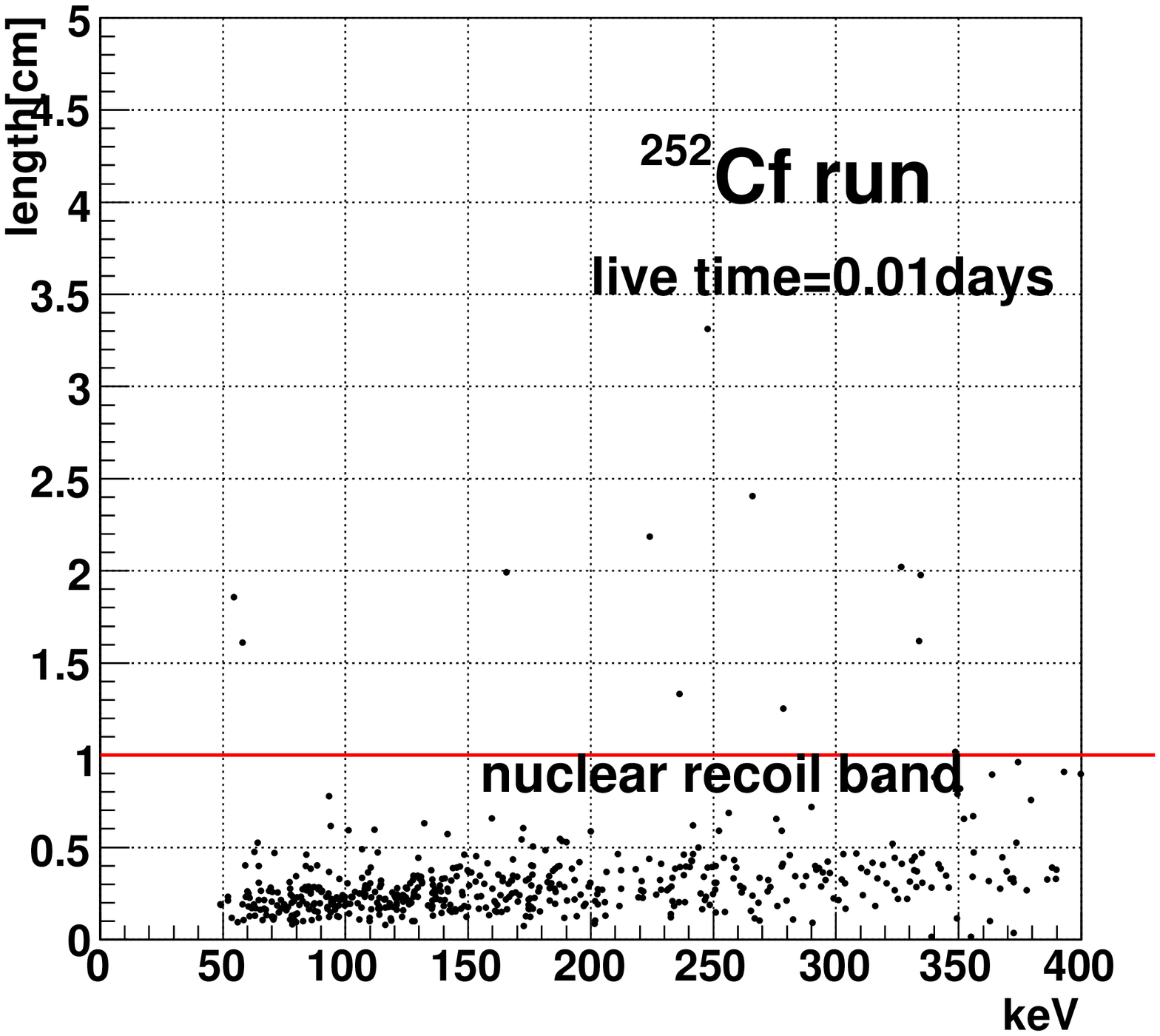}
    \includegraphics[width=0.48\linewidth]{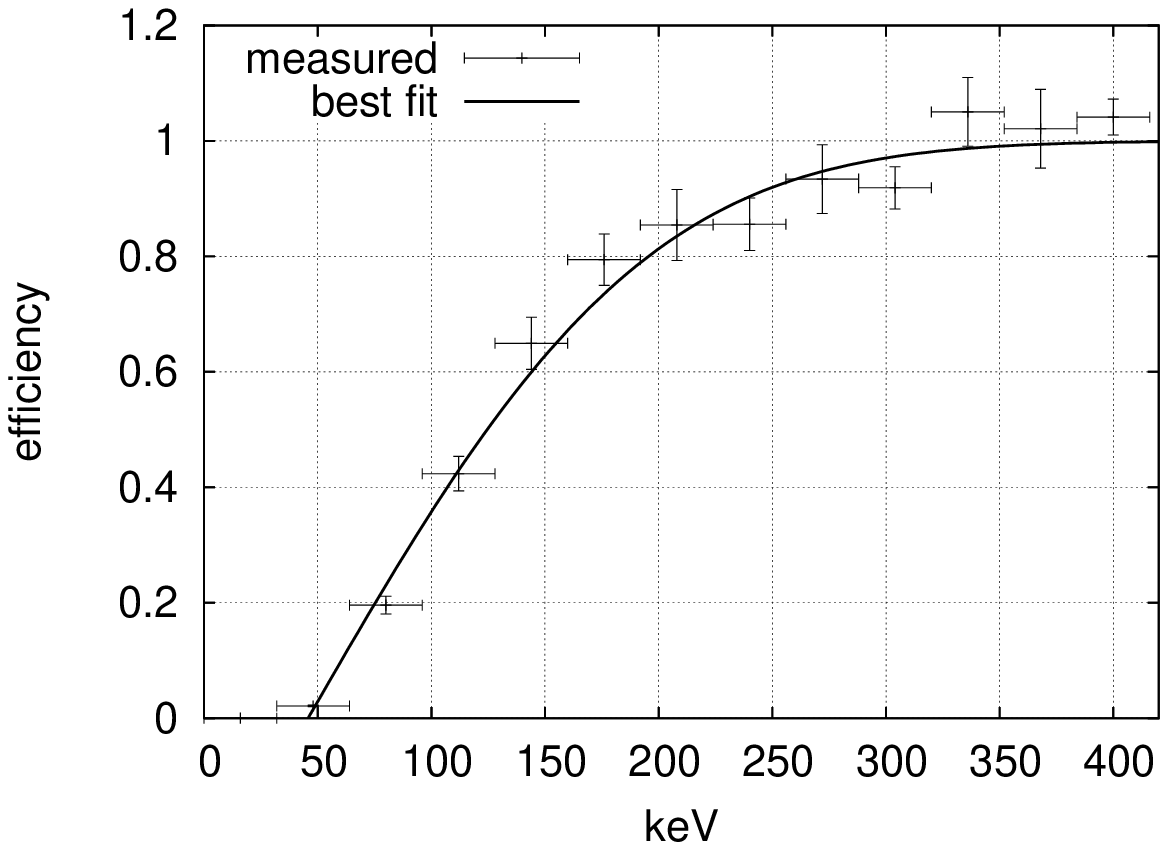}
   \caption{ Measured track-length and energy-deposition correlations with 
a neutron source of $\rm^{252}Cf$ (left) and the detection efficiency dependence on the recoil energy (right). 
}
 \label{fig:scat252Cf}

  \end{center}
\end{figure}

The nuclear-recoil detection efficiency was measured by 
irradiating the $\mu$-TPC with 
fast neutrons 
from a $^{252}$Cf source of  0.72 MBq placed at several postitions. 
We selected the tracks 
within the fiducial volume (21.5 $\times$ 22 $\times$ 31 $\rm cm^3$).
We also required at least three digital hit points for a track.
A typlical track-length and energy-deposition correlation is shown in the 
left panel of Figure~\ref{fig:scat252Cf}. 
A nuclear recoil band is seen along the X-axis.
We set the upper limit of the 
``nuclear recoil band''
at 1 cm.
We compared the spectrum after the selection 
with a simulated one, and 
the calculated ratio was the overall (detection and the selection) 
efficiency.
The efficiency curve is shown in the right panel of Figure~\ref{fig:scat252Cf}.
The efficiency was about 40$\%$ at 100 keV, 
and the efficiency dependence on the recoil 
energy was phenomenologicaly fitted with an error function,  $\epsilon$. 
The best-fit function was $\epsilon=1.0\cdot erf((E-45.8)/165.2)$,
where $erf(x)=\frac{2}{\sqrt{\pi}}\int^{x}_{0}exp(-x^2)dx$ 
is the error function and $E$ is the energy in keV.

\subsection{Gamma-ray rejection factor}
\begin{figure}[h]
  \begin{center}
    \includegraphics[width=0.48\linewidth]{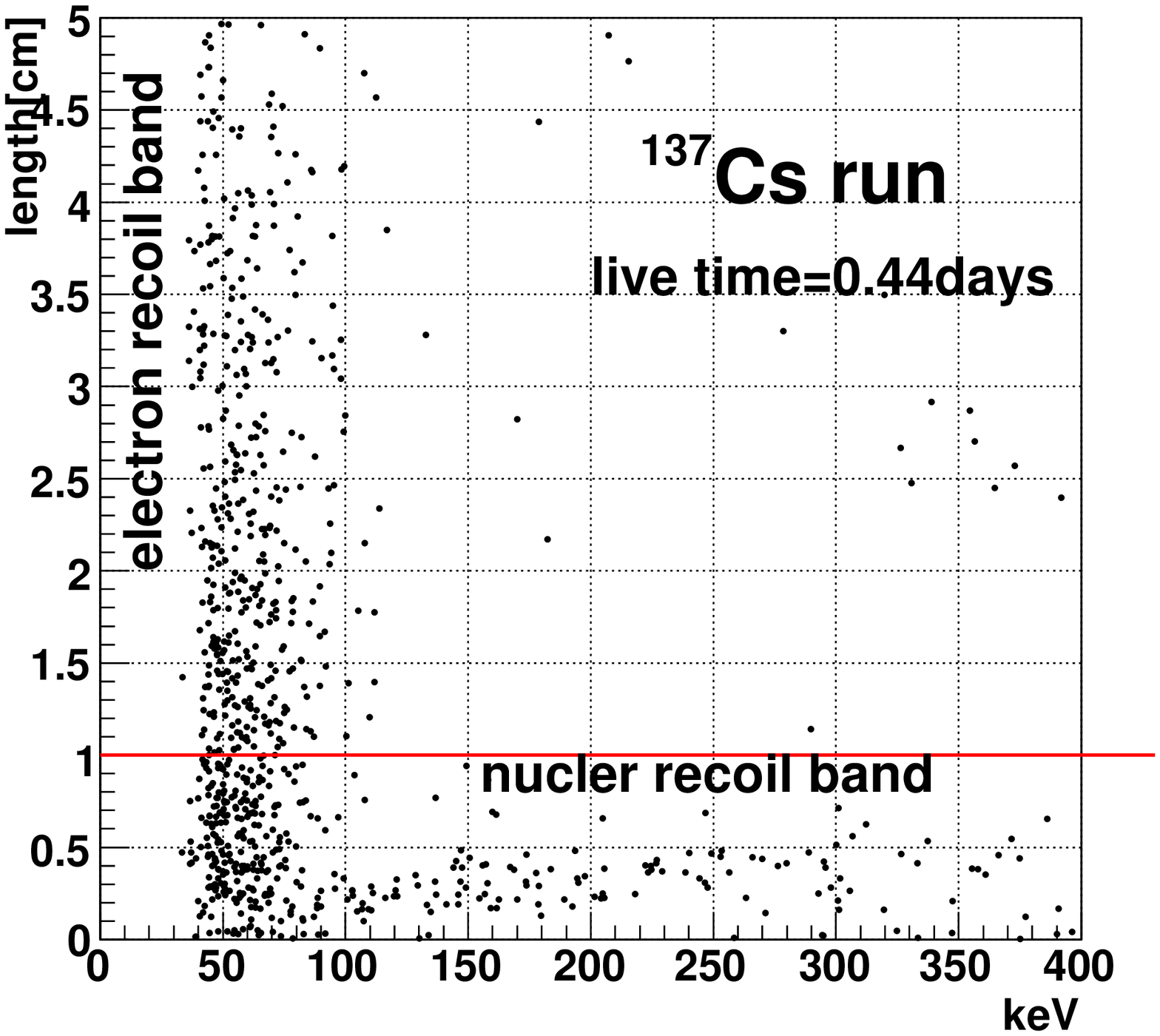}
    \includegraphics[width=0.48\linewidth]{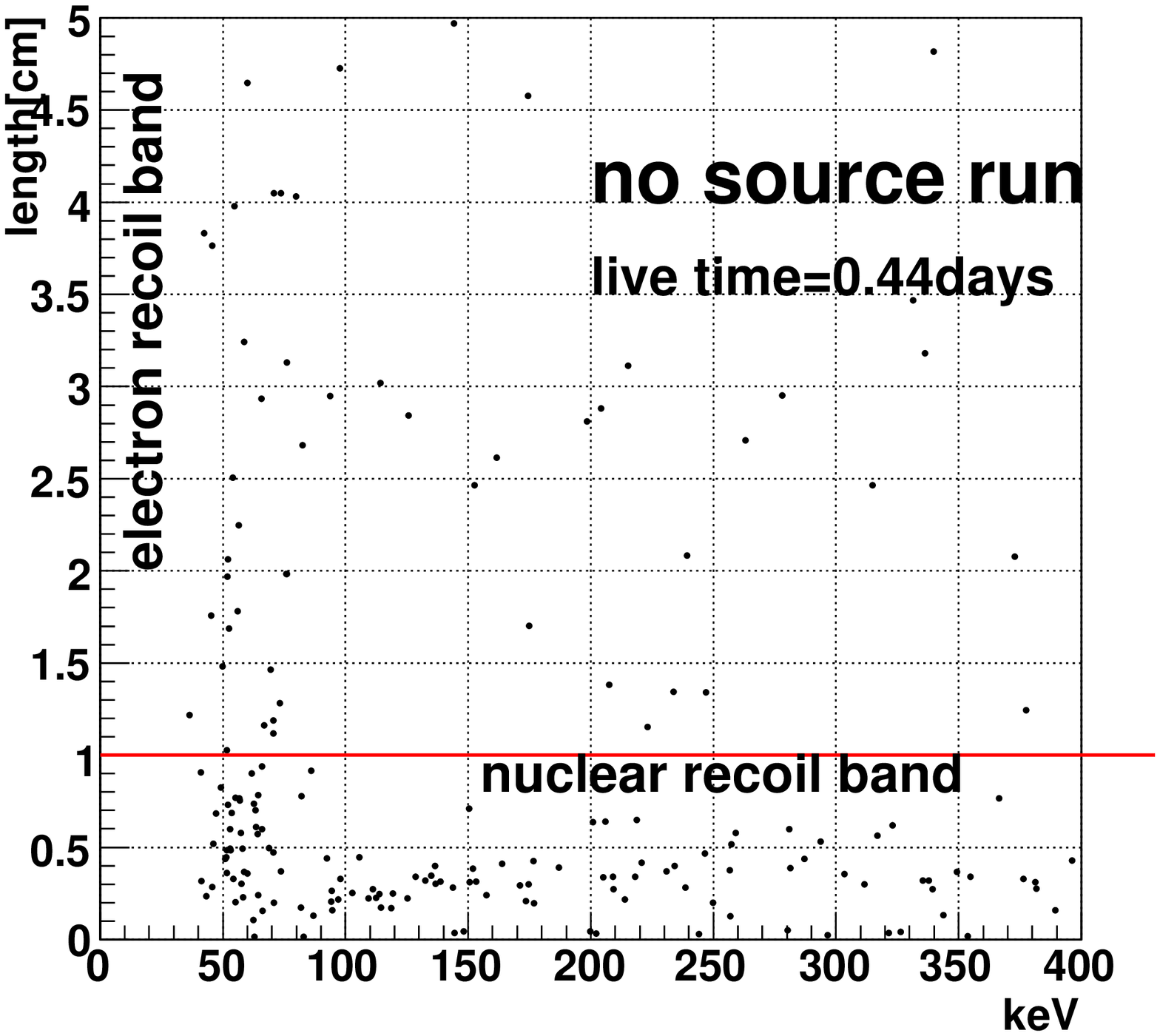}
    \caption{Measured track-length and energy-deposition correlations 
with (left) and without (right) the gamma-ray source of $\rm^{137}Cs$. 
Two bands along the Y- and X-axises are seen in the $\rm^{137}Cs$ run, while the electron recoil band is not seen in the no-source run. 
Several events out of both bands
are thought to be the tracks of protons from the wall of the drift volume.}
    \label{fig:scat137Cs}
  \end{center}
\end{figure}

\begin{figure}[h]
   \begin{center}
\includegraphics[width=0.9\linewidth]{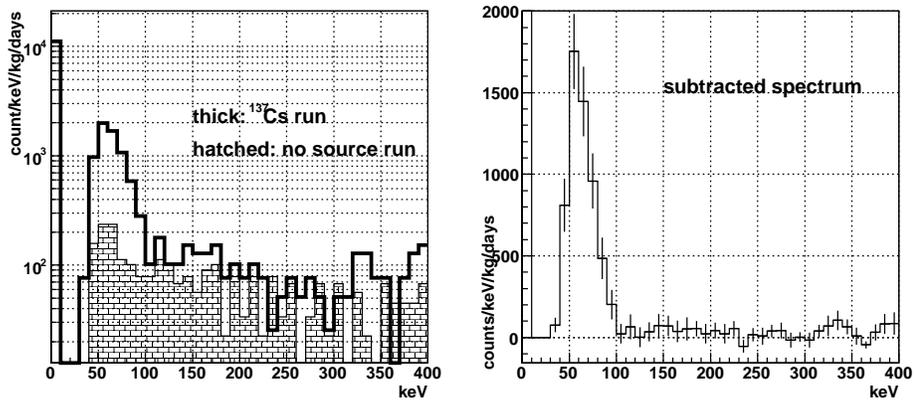}
    \caption{Spectra taken with (thick) and without (hatched) 
the gamma-ray source of $\rm^{137}Cs$ (left) and the subtracted spectrum (right). 
}    \label{fig:spec137Cs}
  \end{center}
   \end{figure}

We demonstrated that 
the track-length and the energy-deposition 
correlation measured with a gaseous TPC 
provided a strong gamma-ray rejection in our previous work\cite{ref:Sekiya_PSD}.
We performed quantitive measurements with a large-volume $\mu$-TPC for a 
WIMP-search measurement in a surface laboratory. 
We irradiated the 
$\mu$-TPC with gamma-rays from a $\rm^{137}Cs$ source of 0.80 MBq, 
placed at 25 cm from the center of the detection volume.
The measured track-length and energy-deposition correlations with and without the 
gamma-ray source are shown in the left and the right panels 
of Figure~\ref{fig:scat137Cs}, respectively.
The live time was 0.44 days for both measurements. 
A band along the Y-axis labeled  ``electron recoil band'' is seen  
only in the results 
with a gamma-ray source, 
which is obviously due to tracks of the recoil electrons.
Another band along the X-axis, seen in both results, is due to 
the tracks of the carbon and fluorine nuclei by the neutron background,  
and are labbeled ``nuclear recoil band''. 
The spectra of these two measurements are shown in the left panel of 
Figure~\ref{fig:spec137Cs}. 
The spectrum of the no-source run was subtracted 
from the that of the $\rm^{137}Cs$ run. The obtained "subtracted''
spectrum is shown in the right panel 
of Fig~ \ref{fig:spec137Cs}. The subtracted spectrum is 
consistent with zero 
in the DM energy range,
within the statistical error.
We then derived the 
gamma-ray rejection factor
by dividing the subtracted spectrum 
by a simulated electron recoil specrum. 
The measured gamma-ray rejection factor in the DM energy ranege was
$(0.57 \pm 0.69 )\times 10 ^{-4}$.
The error is statistical, which is dominated by the background 
in the surface laboratory.
The 90$\%$ C.L. upper limit of the 
gamma-ray rejection factor was $1.5 \times 10 ^{-4}$.

\subsection{Direction-dependent detection efficiency}
\label{section:derection-dependence}
\begin{figure}[h]
   \begin{center}
\includegraphics[width=0.8\linewidth]{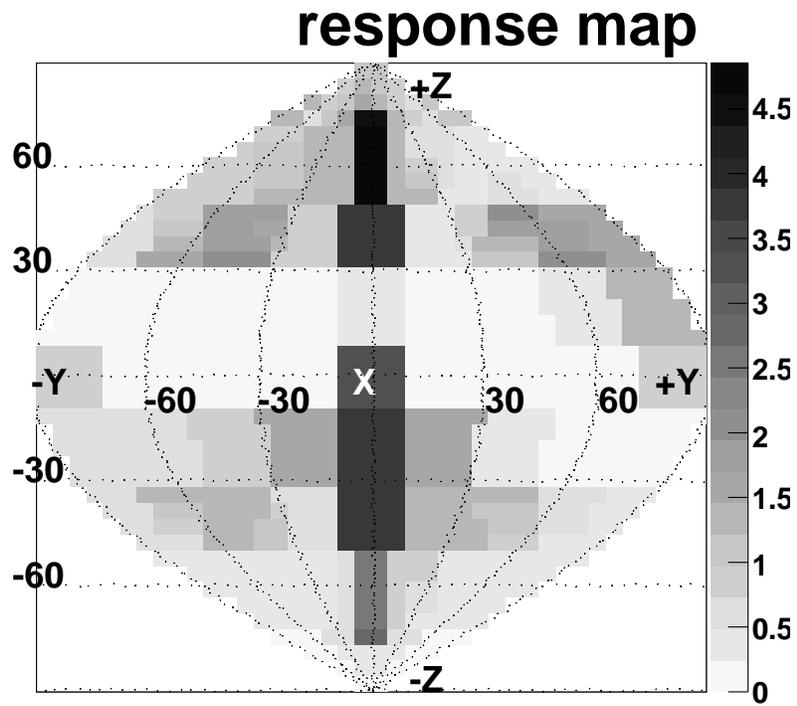}
   \caption{Measured direction-dependent efficiency of the $\mu$-TPC. X means the direction along the X-axis. The map is normalized by the mean efficiency.}
 \label{fig:responsemap}
  \end{center}
   \end{figure}

The $\mu$-PIC has an anisotropic response 
with regard to the three-dimensional track directions.
This non-uniform response is due to
a difference in the detection method of the Z-axis (time information) 
and the other two axes (2-dimensional image taken by the $\mu$-PIC).
We measured this detection efficiency 
dependence on the track direction, or a ''response map'', 
by irradiating the $\mu$-TPC with fast neutrons from several positions
that generated uniform recoils. 
We fitted the digital hit points with a straight line, and 
determined the directions of each nuclear recoil track.
Because we did not measure the track senses in this measurement,   
we fixed the sign of the x component to positive and determined the 
(x,y,z) vector of the track in the coordinates shown in Fig~\ref{fig:TPC}.
The measured response map is shown in Fig~\ref{fig:responsemap}.
The map was normalized by a mean value. 
It can be seen that the correction factor is quite large 
because the fabrication technology of the large-sized $\mu$-PIC 
is still in an R$\&$D phase, and the gain inhomogeniety is still large.
The detector showed 
a relatively high efficiency for tracks in the X-Z and Y-Z planes, but  
showed a relatively low efficiency for tracks in the X-Y plane.

We have not overcome intrinsic weak points of the 
TPC that the efficiency to the tracks parallel to the readout plane 
is relatively low.
It should be noted that the absolute detection efficiency was 
measured by a method described in 
the pervious sections, and this response map was used to 
correct the relative efficiency.


\subsection{Position resolution and angular resolution}
We measured the position resolution using the tracks of the alpha
particles measured in the energy calibration. 
We evaluated the position resolution using the same method as described in  
our previous work \cite{ref:Miuchi_IWORID8}.
The measured position resolution was 0.8 mm~(rms).
The angular resolution of the carbon and fluorine nuclei in 
the DM energy range was estimated in the following manner.
We calculated the sampling pitch of the carbon and fluorine tracks 
from real data.
We then generated the hit points by a simulation 
based on the measured position resolution  and the sampling pitch.
Simulated hit points were fitted with a straight line, and 
the angle between the fitted line and the real direction was calculated.
The distribution of the calculated angle was 
fitted with a Lorentzian and $\gamma=25^{\circ}$, where $\gamma$ is a 
parameter of a Lorentzian that corresponds to HWHM.
We used this value for a direction-sensitive analysis.

\section{Measurement}
\label{section:Measurement}
A dark matter search measurement was performed in a surface 
laboratory at Kyoto University
($Lat.$ $35^{\circ} 2'$ N, $Long.$ $135^{\circ} 47'$ E, 
third floor in a five-story building).
The measurement was performed from November 1st, 2006 to 
November 27th, 2006.
We set the $\mu$-PIC plane horizontally and 
re-aligned the X- and Y-axis of the $\mu$-TPC twice 
(three different $\phi$ indicated in Figure~\ref{fig:TPC}) 
so as to cancel any remaining static errors.
Table \ref{table:TPCoperation} lists the  measurement properties.
The total live time was 16.71 days and 
the exposure was 0.151 kg$\cdot$ days.
The sum of the calibration and maintenance time was 7.51 days,
and the dead time due to the data acquisition was 1.0 days.


The $\mu$-TPC was filled with 150 Torr $\rm CF_4$, 
and the vacuum chamber was sealed without any gas circulation..
The drift velocity, total gas gain, and the energy resolution 
were monitored every two or three days during the measurement.
The drift velocity was 7.6~cm/$\mu$s at the beginning of the 
measurement and decreased to 
6.9~cm/$\mu$s at the end. 
This decrease is thought to be due to the 
out-gas from the detector components. 
We used the measured velocities for the analysis.
The mean gas gain in the entire detection area was 2300, and 
a time-dependent variation was not observed 
within a measurement error of 4$\rm\%$ (rms).
The energy resolution was also stable within a 
measurement error of 7$\rm\%$ (rms).
The time-dependence was small compared to the energy resolution, itself, 
and the angular resolution in this measurement.


\begin{table}
\begin{center}
\begin{tabular}{l l l}
X direction($\psi$) & source &live time [days] \\ \hline
$-5^{\circ}$ & - & 4.37 \\ \hline
$40^{\circ}$ & - & 5.25 \\ \hline
$18^{\circ}$ & - & 4.95 \\ \hline
$18^{\circ}$ & $^{137}$Cs & 2.14 \\ \hline
\end{tabular}
\caption{Properties of the dark matter measurements.}
\label{table:TPCoperation}
\end{center}
\end{table}

\section{Results}
\label{section:Results}
\begin{figure}[h]
   \begin{center}
\includegraphics[width=0.9\linewidth]{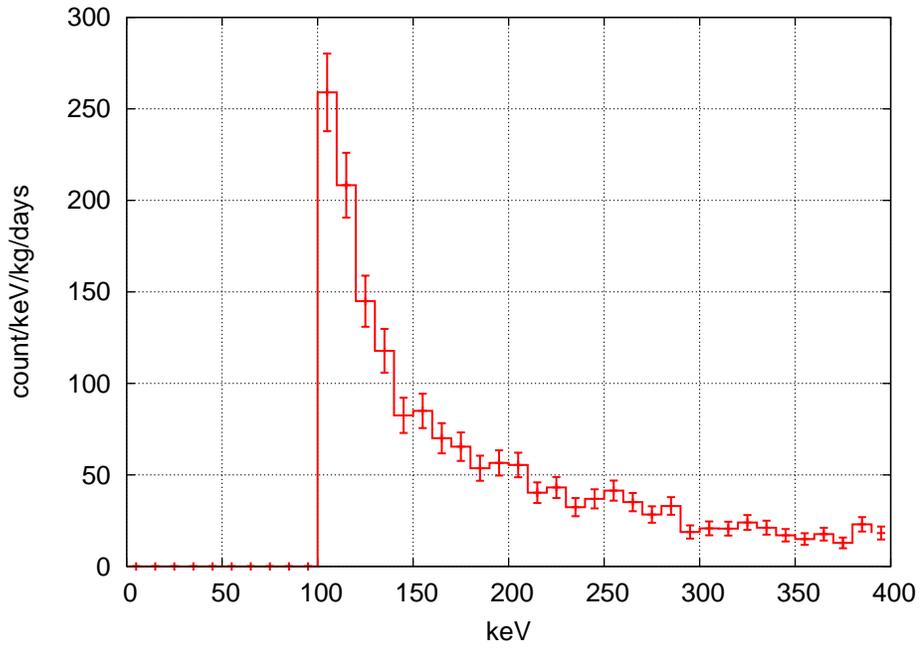}
   \caption{Measured energy spectrum. The detection efficiency is taken into account.}
 \label{fig:spec}
  \end{center}
   \end{figure}

\begin{figure}[h]
   \begin{center}
\includegraphics[width=1.\linewidth]{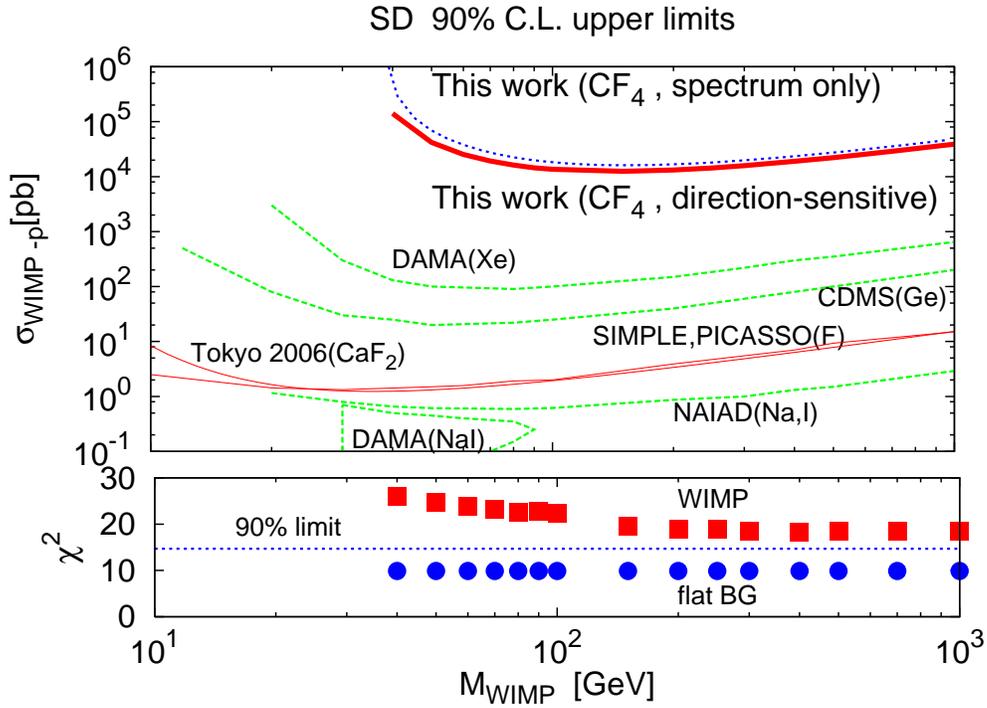}
   \caption{90$\%$ C.L. upper limits on the WIMP-proton spin-dependent 
cross section (upper) and $\chi^2$ values (lower) as functions 
 of the WIMP mass.
The thick solid and dotted lines show the limits obtained with and without the direction information, respectively. 
Limits from other experiments 
( DAMA(Xe)\cite{ref:DAMA_Xe}, DAMA(NaI)\cite{ref:DAMA_annual}, NAIAD\cite{ref:NAIAD}, Tokyo $\rm CaF_2$\cite{ref:Tokyo_CaF2}, SIMPLE 2005\cite{ref:SIMPLE2005}, PICASSO\cite{ref:PICASSO}, and CDMS\cite{ref:CDMS2}) are shown for comparison. 
The filled squares show the $\chi^2$ minimum values of the best-fit WIMP 
distribution, 
the filled-circles show
the best-fit flat $\cos \theta$ distribution, and  
the dotted line show the $\chi^2$ values at the 90$\%$ C.L. upper limit. }
 \label{fig:limit}
  \end{center}
   \end{figure}

The spectra did not show a statistically significant difference among the
four conditions due to the strong gamma-ray rejection.
We therefore combined the spectra obtained in the four measurements, 
including even the spectrum obtained in the  
gamma-ray rejection measurement with a $^{137}$Cs source.
Figure~\ref{fig:spec} shows
the obtained spectrum in the DM energy range.
The spectrum can be explained by 
a typical neutron flux in a 
surface laboratory of $O$(1$\rm \times10^{-2} n\cdot cm^{-2}\cdot s^{-1}$).
We initially derived the limits of the WIMP-proton spin-dependent cross section 
only from the spectrum (conventional method).
We used the parameters given in Table \ref{table:app and nu params}, and 
followed Ref.\cite{ref:LewinSmith} for the calculation.
The thin-dotted line, which is indicated by ''spectrum only'', 
in Figure~\ref{fig:limit} shows the exclusion limits.

\begin{table}
\begin{center}
\begin{tabular}{l l}
\hline
WIMP velocity distribution & Maxwellian  \\ \hline
solar velocity & $v_{\rm s}$=244km$\cdot \rm {s}^{-1}$ \\ \hline
Maxwellian velocity dispersion& $v_{0}$=220km$\cdot \rm {s}^{-1}$ \\ \hline
escape velocity &$v_{\rm esc}$ =650 km$\cdot \rm s^{-1}$ \\ \hline
local halo density & $\rm 0.3\,GeV\cdot cm^{-3}$ \\ \hline
spin factor of $\rm {}^{19}F$ & $\lambda^2J(J+1)=0.647$ \\ \hline \hline
\end{tabular}
\caption{Astrophysical and nuclear parameters used to derive the WIMP-proton limits. }
\label{table:app and nu params}
\end{center}
\end{table}

\begin{figure}[h]
   \begin{center}
\includegraphics[width=0.8\linewidth]{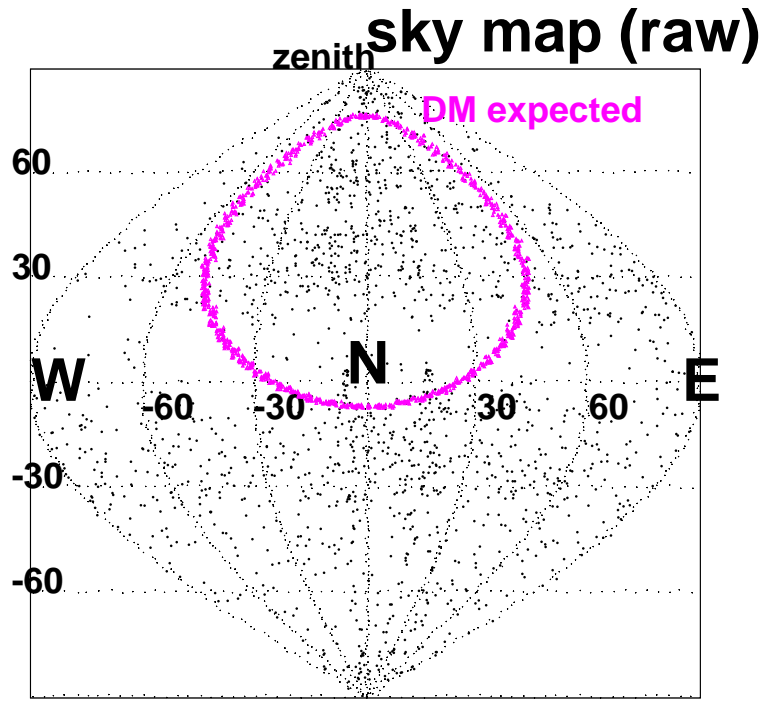}
   \caption{Obtained sky map of the north sky. The recoil directions of the 
fluorine and carbon nuclei are used. A raw map and a corrected map by the relative 
efficiency dependent on the directions are shown.
The thick line indicates the direction of the solar motion (direction of the WIMP-wind).}
 \label{fig:skymap}
  \end{center}
   \end{figure}

\begin{figure}[h]
   \begin{center}
\includegraphics[width=1.0\linewidth]{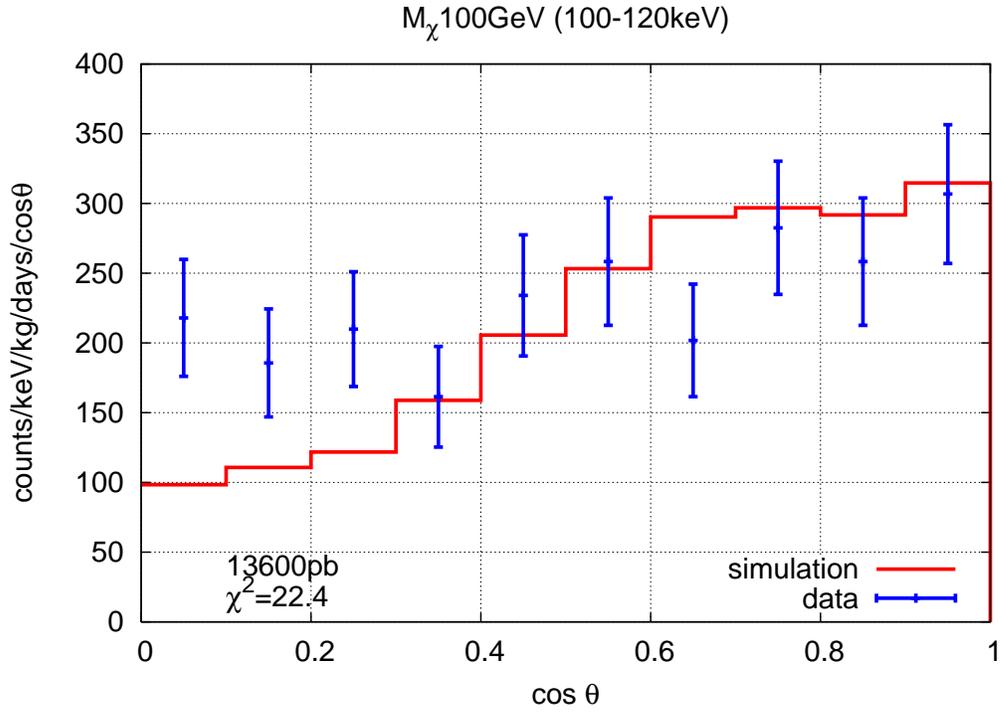}
   \caption{Measured (with error bars) and expected (histogram) 
distribution of the angle between the recoil 
direction and the WIMP direction. 
The expected histogram is 
that with $M_\chi$=100GeV, 
100--120 keV bin, and 1.36$\times10^4$pb.}
 \label{fig:costheta}
  \end{center}
   \end{figure}

We next performed a direction-sensitive analysis.
We created a sky map of the north sky with 
the direction of 
the three-dimensional nuclear recoil tracks.
Because we did not detect the sense of each track, 
the map was restricted to half of the sky larger
and the southern part of the sky was folded.
We plotted every nuclear event (1686 events in  a 0.151 kg$\cdot$days exposure)
in Figure~\ref{fig:skymap}.
The larger mark, labled "DM expected'', 
is the direction of the solar motion or the Cygnus 
from which the WIMP-wind is expected.
The WIMP-wind direction has a diurnal motion around the polestar. 
We calculated $\theta$, the angle between the WIMP-wind direction 
and the recoil direction for each event, and drew the  
the $\cos \theta$ distribution shown in
Figure~\ref{fig:costheta}.
We compared this measured 
$\cos \theta$ distribution with the expected WIMP-signal.
The expected WIMP-signal was prepared by the following method.
We first created a $\cos \theta$ distribution for a 
given WIMP mass and a recoil energy while considering the energy resolution. 
The DM energy range was divided into 15 bins, and 
the energy and angular resolution of the detector were taken into account.
We used $70\%$ (FWHM) for the energy resolution and an angular resolution of 
$\gamma=25^{\circ}$.
We followed Ref.~\cite{ref:Spergel_WIMP} 
for the angular distribution and Ref.\cite{ref:LewinSmith} for the count rate.
We then created a sky-map expected by the WIMP events 
with this $\cos \theta$ distribution while considering 
the derection-dependence(Figure~\ref{fig:responsemap}) and the angular resolution.
We used this sky map to create the expected 
$\cos \theta$ distribution 
for a given WIMP mass and energy bin.
Figure~\ref{fig:costheta} shows one of these $\cos \theta$ distributions.
Here, the data was fitted with the distribution for $M_\chi$=100GeV and 
100--120 keV bin, and the cross section that gave the minimum $\chi^2$ 
was 1.36 $\times$ $10^4$pb and $\chi^2/d.o.f=22.4/9$. 
This best-fitted WIMP signal was rejected at the 90$\%$ confidence level 
by $\chi^2$ tests. 
The cross section that would give the minimum $\chi^2$ was calculated for 
each energy bin, and the smallest cross section was taken as the limit 
for the given WIMP mass.
The upper and lower panels  
in Figure~\ref{fig:limit} show 
the cross section limits and the corresponding $\chi^2$ values 
as a function of the WIMP mass, respectively.
The limits are shown with a thick-solid line (labeled ''direction-sensitive'') 
and the $\chi^2$ values are shown by filled-square marks.
The $\chi^2$ values at the 90$\%$ C.L. upper limit are shown by a dotted line
in the lower panel in Figure~\ref{fig:limit}.
The WIMP signals were rejected at the 90$\%$ confidence level
because the best-fit $\chi^2$ values were all above the 90$\%$-$\chi^2$ line.
The data was also fitted with a flat $\cos \theta$ distribution and 
$\chi^2/d.o.f=9.9/9$ was obtained independent of the WIMP mass. 
The flat $\cos \theta$ was not rejected at the 90$\%$ confidence level.
It was demonstrated that the 
direction-sensitive method 
gave a limit slightly better than 
the conventional 
spectrum-only method.
The difference between this result and other experiments is explained by the 
neutron background in the surface laboratory.
This is the first limit set by 
a dark matter search experiment with a gaseous tracking device. 
As for the spin-dependent limit, this is the first result 
by a direction-sensitive experiment.
It should be stressed that 
the dark matter search experiment with a gaseous tracking device 
can actually set a limit on the WIMPs.
These results are 
very convincing and encouraging 
towards developing a future WIMP-wind detection with large-volume detectors.







\section{Discussions}
\label{section:discussion}
We have shown that WIMP-proton cross section 
limits can be obtained by a direction-sensitive method. 
However, much work is necessary to reach
the sensitivity achieved by other experiments performed with  
solid and liquid detectors.
The sensistivity is restricted by the neutron background in a 
surface run, therefore, we are going to perform 
a dark matter search experiment
in an underground laboratory. 
The fast neutron flux in Kamioka Observatory 
((1.9 $\pm$ 0.21) $\rm \times 10^{-6} n\cdot cm^{-2}\cdot s^{-1}$) is 
more than 3 orders of magnitude smaller than a typical fast neutron flux in a 
surface laboratory.
It it expected that the background sources, 
such as radioactive isotopes in the detector components, would 
limit the sensitivity in the first stage of an underground run.
We are going to replace the detector components with low-activity materials.
The angular resolution is not good enough to 
take full advantage of the direction-sensitive method over
the ``spestrum only'' method.
We are going to 
investigate the detector response using a neutron source of a known energy. 
We plan to study the operation of the $\mu$-TPC at a lower pressure of 30 Torr
to lower the energy threshold.
We are going to develop a larger readout area by placing many pieces of the 
of $\mu$-PICs of 31 $\times$ 31 $\rm cm^2$.  
After these technological breakthroughs are achieved, 
we expect to reach the SUSY predictions 
with an exposure of more than one hundred kg$\cdot$days, which 
we have calculated in our previous work\cite{ref:Miuchi_PLB}.

\section{Conclusion}
\label{section:Conclusion}
We developed a three-dimensional tracking device, 
and performed a direction-sensitive 
dark matter search measurement in a surface laboratory.
By using 150 Torr carbon-tetrafluoride ($\rm CF_4$) gas, 
we set the first limit on the spin-dependent WIMP-proton cross section 
by a direction-sensitive method.
Thus, we have demonstrated that  
a dark matter search experiment with a gaseous tracking device 
can actually set a limit on WIMPs.
Furthermore, we have demonstrated that this method will potentially play a 
certain role in revealing the nature of dark matter 
when a low-background large-volume detector is developed. 

\section*{Acknowledgements}
This work was supported by a Grant-in-Aid in Scientific Research of 
the Ministry of Education, Culture, Sports, Science, Technology of Japan; 
research-aid program of Toray Science Foundation;
and Grant-in-Aid for the 21st Century COE 
''Center for Diversity and Universality in Physics''.

\newpage

\newpage

\newpage
\newpage

\newpage

\newpage

\newpage

\newpage

\newpage

\end{document}